\documentclass[
    twocolumn,
    floatfix
]{aastex631}
\usepackage{hyperref}
\usepackage{color}

\newcommand{\msun}{{\rm M_\odot}}
\newcommand{\eg}{e.\,g.}

\newcommand{\action}[1]{{\color{teal} #1}}

\shorttitle{The Era of Binary Supermassive Black Holes}

\begin{document}

\title{The Era of Binary Supermassive Black Holes:\\ Coordination of Nanohertz-Frequency Gravitational-Wave Follow-up}

\author[0000-0003-4052-7838]{S. Burke-Spolaor}
\affil{Department of Physics and Astronomy, West Virginia University, P.O. Box 6315, Morgantown, WV 26506, USA}
\affil{Center for Gravitational Waves \& Cosmology, West Virginia University, Chestnut Ridge Research Bldg., Morgantown, WV 26505, USA}
\affiliation{Department of Physics and Astronomy, Johns Hopkins University,
3400 North Charles Street, Baltimore, Maryland 21218, USA}
\email{sarah.spolaor@mail.wvu.edu}
\author[0000-0002-7835-7814]{T. Bogdanović}
\affil{School of Physics, Georgia Institute of Technology, 837 State St NW, Atlanta, GA 30332, USA}
\affil{Center for Relativistic Astrophysics, Georgia Institute of Technology, Atlanta, GA 30332, USA}
\author[0000-0002-1271-6247]{D. J. D'Orazio}
\affiliation{Space Telescope Science Institute, 3700 San Martin Drive, Baltimore , MD 21}
\affiliation{Department of Physics and Astronomy, Johns Hopkins University,
3400 North Charles Street, Baltimore, Maryland 21218, USA}
\affiliation{Niels Bohr International Academy, Niels Bohr Institute, Blegdamsvej 17, DK-2100 Copenhagen Ø, Denmark}
\author[0000-0002-3719-940X]{M. Eracleous} 
\affil{Department of Astronomy \& Astrophysics, Penn State University, 525 Davey Lab, 251 Pollock Road, University Park, PA 16802}
\author[0000-0003-3703-5154]{S. Gezari}
\affil{Department of Astronomy, University of Maryland, College Park, MD  20742-2421}
\author[0000-0002-3168-0139]{M. J. Graham} 
\affil{Division of Physics, Maths and Astronomy, California Institute of Technology, 1200 E California Blvd, Pasadena, CA 91125, USA}
\author[0000-0002-1146-0198]{K. G\"{u}ltekin}
\affil{University of Michigan, Department of Astronomy, 1085 S University Ave, Ann Arbor, MI 48109, USA}
\author[0000-0003-2742-3321]{J. Hazboun}
\affil{Department of Physics, Oregon State University, Corvallis, OR 97331, USA}
\author[0000-0002-4307-1322]{C. M. F. Mingarelli}
\affil{Department of Physics, Yale University, New Haven, CT 06520, USA}
\author[0000-0001-6022-0484]{G. Narayan}
\affil{University of Illinois, Urbana-Champaign, 1002 W. Green St., Urbana, IL 61801, USA}
\affil{NSF Simons SkAI Institute, 875 N. Michigan Ave., Chicago, IL 60611, USA}
\author[0000-0001-5681-4319]{P. Petrov}
\affil{Department of Physics and Astronomy, Vanderbilt University, 2301 Vanderbilt Place, Nashville, TN 37235, USA}
\author[0000-0001-7678-8218]{N. Veronesi}
\affiliation{Department of Physics and Astronomy, Washington State University, 1245 Webster Hall, Pullman, WA 99164, USA}




\begin{abstract}
%

Here we summarize discussions and conclusions from the conference ``The Era of Binary Supermassive Black Holes: Coordination of Nanohertz-Frequency Gravitational-Wave Follow-up,'' held at the Aspen Center for Physics from February 2-7, 2025. The meeting facilitated a crucial knowledge exchange between electromagnetic and gravitational-wave theorists, observers, and cyber-infrastructure experts. The central goal was to guide the development of multi-messenger follow-up strategies for binary supermassive black hole detections by pulsar timing arrays. To build a common basis of understanding for the broader scientific community, this summary outlines the main considerations and recommendations from the meeting, summarizes the knowledge gaps identified, and ends with a potential roadmap to catalyze discussion about the search for electromagnetic counterparts to massive black hole binaries detected by pulsar timing arrays.
\end{abstract}

\section{Conference Format}
The central premise of the conference was to discuss topics relevant to the following question: ``What is needed to take us from a PTA report of a CW detection to a confident MM identification?''\footnote{\noindent {\bf Acronyms used throughout this document:}\\
AGN = Active Galactic Nucleus\\
CW = Continuous Wave (sinusoidal or cyclic signal from MBHB)\\
EM = Electromagnetic\\
GW = Gravitational Wave\\
GWB = Gravitational-Wave Background\\
IPTA = International Pulsar Timing Array\\
LIGO = Laser Interferometer Gravitational-Wave Observatory\\
LISA = Laser Interferometer Space Antenna\\
PTA = Pulsar Timing Array\\
MBH = Massive Black Hole\\
MBHB = Massive Black Hole Binary\\
MM = Multi-messenger\\
MMA = Multi-messenger astrophysics\\
MPTA = MeerKAT Pulsar Timing Array\\
NANOGrav = Nanohertz Observatory for Gravitational Waves\\
SCIMMA = Scalable Cyber-Infrastructure to support MMA}

Conference mornings were divided into thematic areas with broad plenary talks to build participants’ awareness of each topic.
The thematic areas were divided into sessions enumerated on the schedule as follows: 1) PTA status and observing capabilities, 2) Implications of the GWB for MBHB populations and physics, 3) MBHB emission theory, 4) MBHB observations and ambiguities, 5) Capabilities of future electromagnetic observatories, and 6/7) Cyber-infrastructure needs of PTA-related MMA coordination. Posters and contributed talks were also presented.

Twelve interactive parallel sessions took place in the evenings of the conference. During these sessions, moderators and participants discussed specific questions and identified action items. 
The questions generally fell into four themes: 1) comparing theory and observations; 2) designing follow-up strategies; 3) discussing EM-GW coordination needs; and 4) designing MBHB detection surveys.
These discussions each lasted one hour, had two moderators, and were attended by 10--20 participants. After the parallel discussions, all participants convened to share and discuss their conclusions.

The schedule with links to session recordings is publicly available,\footnote{\url{https://tinyurl.com/nano-aspen2025-schedule}} as are presenter slide decks.\footnote{\url{https://tinyurl.com/nano-aspen2025-slides}}

\section{PTAs and MBHBs: Common Assumptions / Preliminaries}\label{sec:preliminaries}
Many discussions were built from a set of common assumptions about PTAs and their target populations.

For readers less familiar with PTA analysis, it will help to see the talks by Witt and Vigeland for a broad introduction to PTAs and CW detection techniques, and the ``Pulsar Q\&A'' session on Wednesday Feb.~5, in which PTA astronomers clarified many points asked by conference attendees about PTA analysis.

Here we identify some assumptions about PTAs and MBHB populations that guided meeting discussions.

{\bf MBHBs form the dominant signal in the GWB:} PTAs have shown evidence of a GWB signal consistent with numerous potential astrophysical interpretations \citep{nano-15yr-evidence,ppta-gwb,epta-gwb,cpta}, and MBHBs as the source of the GWB evidence are promising but not definitive. In most talks in this conference, it was assumed for the estimation of CW rates, properties, and time-to-detection that the GWB emerging in pulsar timing data sets is dominated by GWs from MBHBs. 

{\bf The CW signal-to-noise ratio will rise gradually:} If the previous assumption holds, CWs may already be present in PTA data at moderate to very low (sub-detection) thresholds. 
PTA detections of CWs will evolve to gradually higher signal strength. That is, they are unlikely to be suddenly identified with high significance in the data, similar to how the GWB appears to be emerging as a signal of growing strength.

\begin{figure}
    \centering
    \includegraphics[width=1.0\columnwidth]{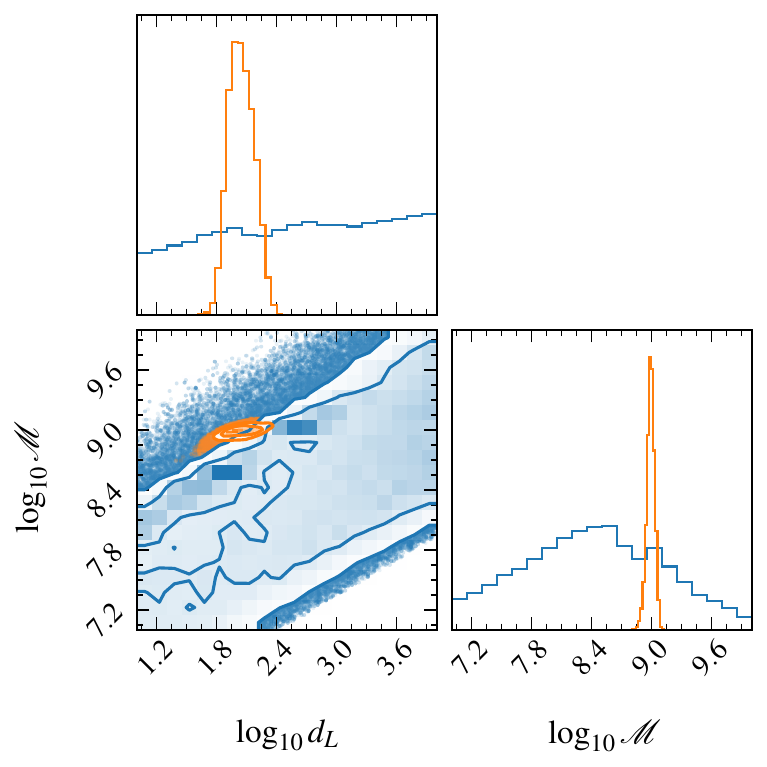}
    \caption{Talks by Witt and Petrov highlighted that in a PTA-based CW detection, mass and distance will be covariant until high signal-to-noise is reached. The above plot by P.~Petrov is an illustrative example based on simulations described in \cite{petrov+24}. Shading represents the density of posterior samples in the parameter space. The colored contours represent the 68\% and 95\% credible intervals for two simulated MBHB CW detections. Scatter points outside the contours are posterior samples that fall outside of the 95\% credible interval. Both simulations used $\mathcal{M}=10^9~\msun$ and $f_{\rm gw}=20$~nHz in an ``IPTA Data Release 3''-like 116-pulsar PTA. The distance of the source was varied to induce a CW with a theoretical signal-to-noise ratio of 8 (blue, corresponding to 303~Mpc) and 16 (orange, 151.5~Mpc).
    }
    \label{fig:banana}
\end{figure}

{\bf Source properties measured by PTAs:} 
At low detection significance, PTAs should be able to provide the following information:
\begin{itemize}
    \vspace{-1.75mm}\item GW frequency $f_{\rm gw}$ to good (sub-dex) accuracy. 
    \vspace{-1.75mm}\item Right ascension and declination, at first poorly constrained (hundreds of square degrees or more \citealt{goldstein+19,petrov+24}).
    \vspace{-1.75mm}\item A covariant posterior on chirp mass $\mathcal{M}$ and luminosity distance $D_{\rm L}$. The covariance forms an elongated shape maximized along a line running between massive, distant systems and less massive, nearby systems (illustrated example in Fig.~\ref{fig:banana}). 
\end{itemize}
\vspace{-1.5mm}
Chirp mass and source distance will be covariant because decoupling those terms requires the detection of frequency evolution, which is possible but unlikely for PTAs to achieve in the low-significance regime. Circular orbits can be considered as close to stable (no meaningful chirp) when the orbital inspiral timescale is long compared to the PTA timing baseline, $T\lesssim\tau_{\rm gw}$ where
\begin{equation}
\hspace{-2mm}\tau_{\rm gw}\simeq321.1\,{\rm yr}\bigg(\frac{P_{\rm orb}}{{\rm 1~yr}}\bigg)^{8/3}\bigg(\frac{10^9\,\msun}{M_{\rm tot}}\bigg)^{2/3}\,\bigg(\frac{10^9~\msun}{\mu}\bigg)
\end{equation}
Here $M_{\rm tot}$ is the total mass of the binary, $\mu$ is reduced mass and $P_{\rm orb}$ is the source-frame orbital period (c.f. \citealt{maggiore}). Millisecond pulsars were discovered in only 1982 and IPTA data sets span $T\simeq20$--$30$~years, thus PTAs will not fit this condition in our lifetimes for the coherent, quadrupolar GW modulation of Earth's space-time (``Earth term''). Thus, the degeneracy may only be broken when PTAs have a bright enough detection to significantly detect the pulsar-local space-time modulation (``pulsar-term,'' see talk by C. Witt), which encodes a long-timescale MBHB evolution signal.

{\bf Likely mass, distance, and GW frequency of CW detections:} Based on cosmological simulations, we generally expect a horizon of PTA CW detection of around 1.5 Gpc ($z\lesssim0.4$) over the coming two decades \citep[e.g.]{gardiner+24}. CW emission is only formally detectable when it is resolvable from the background signal.
Maximizing individual source GW strain requires maximizing both $M_{\rm tot}$ and mass ratio $q$, therefore CW detections should preferentially arise from major, equal-mass galaxy mergers of the most massive galaxies ($q\sim1$, ${\rm log}(M_{\rm BH}/{\rm M_\odot}) \gtrsim 8.5$ hosts). Population simulations generally support these assumptions \citep[\eg][]{gardiner+24}.



\section{Main Outcomes}
Here we aim to capture the salient discussion points, action items, and recommendations identified during the conference.
\action{\bf The colored text below indicates points where new action is needed to make progress}; in some cases, work on these points is already underway.
The summaries here represent an attempt of the authors to provide a sense of the concensus derived from discussions, but do not necessarily represent a unified viewpoint of all participants.


\subsection{(Un)certainty of Direct Emission Signatures}\label{sec:em}
Electromagnetic emission from MBHBs was a main point of discussion for Sessions 3 and 4, and many evening discussions.

The broad consensus is that single MBHs in an accretion environment are also capable of generating most proposed binary signatures. A poorly localized CW will have many potential host galaxies. Thus, we will need multi-method observation on many targets to compare the likelihood of association with the CW; proof will be built based on evidence from numerous independent physical processes within a MBHB environment \citep[\eg][]{doraziocharisi}. For instance: broad-band SEDs and flux monitoring may probe circumbinary disk models, spectra can probe MBH-associated gas dynamics, GHz radio jet imaging can probe binary-induced precession; meanwhile, time-domain monitoring can quantify periodic behaviors. All of these methods should provide a model with a self-consistent interpretation, and one that agrees with the CW parameterization. 

A crucial part of this topic is the accurate modeling of MBHB nuclear environments. Talks in Sessions 3 (Tuesday: Miller, D'Orazio) and 7 (Thursday: Noble, Siwek) and several posters 
described binary accretion simulations, focusing on binary orbital evolution and associated EM signatures. The presentations focused on two predominant approaches to the problem: 1) those that include more of the physics necessary to describe a realistic accretion flow and its EM properties (combinations of general relativity, magneto-hydrodynamics, and radiation hydrodynamics) and 2) those with higher resolution and longer run times but that only include hydrodynamics (and sometimes simplified cooling functions, often in 2D).  
It was clear that \action{further connection between these approaches is needed to solve this difficult problem. For example: can controlled comparisons (similar disk and binary properties) help to understand robust/non-robust features? And, can insights from the more costly simulations help speed up parameter-space exploration or improve predictions for observable signals from accreting binaries?}

\begin{figure}
    \centering
    \includegraphics[width=1.0\columnwidth,trim=0mm 0mm 1mm 0mm,clip,]{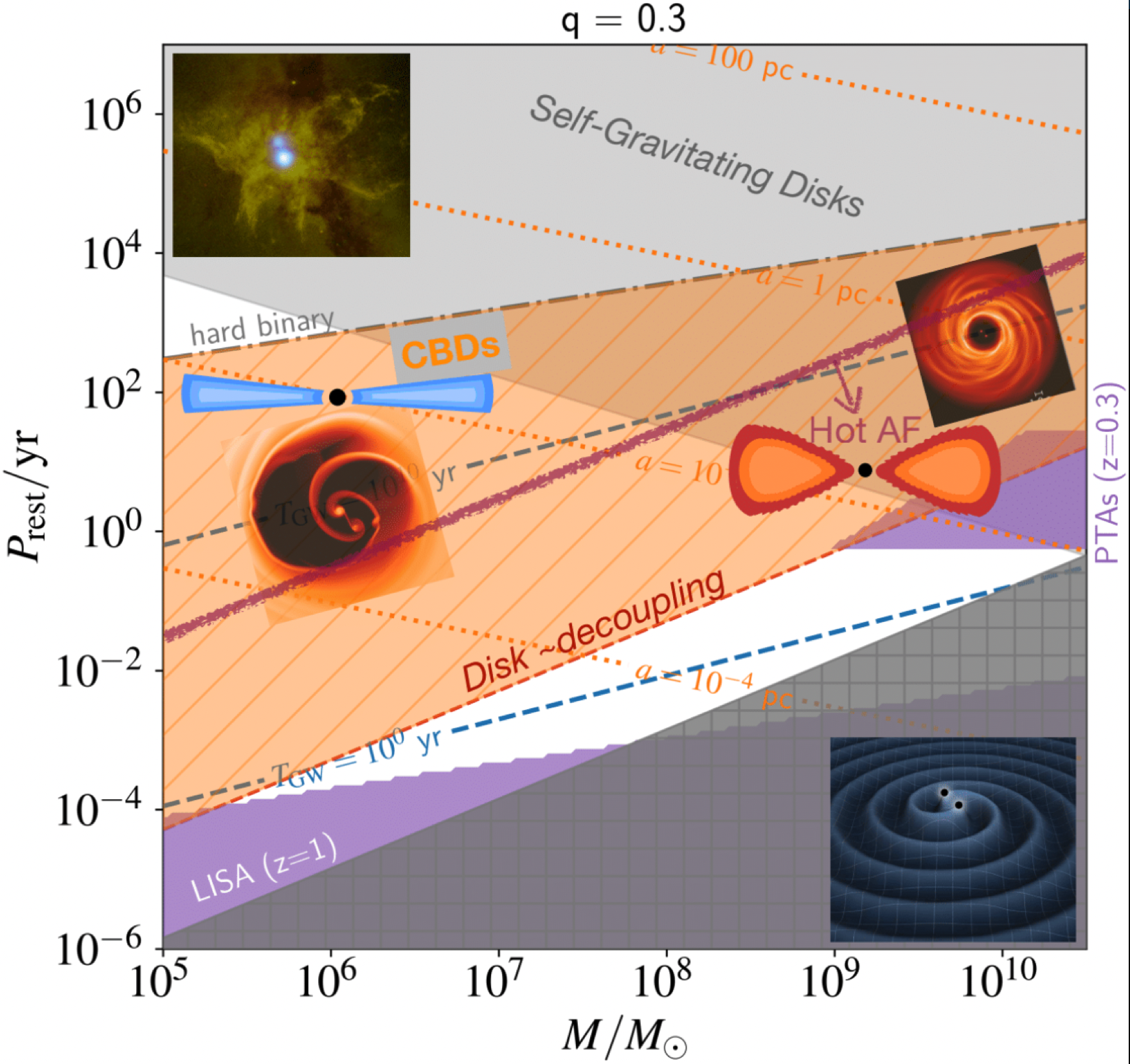}
    \caption{A schematic representing regimes of gas disk (thermo)dynamics around a circular, $q=0.3$ MBHB of rest-frame orbital period $P_{\rm rest}$ and total mass $M$, based on physical arguments outlined in the talk by D'Orazio.
    Binaries within the regime between the ``hard binary'' line (below which MBHB orbital energy exceeds stellar dynamical energy) and the ``disk decoupling'' line (below which the binary's GW evolution may drive binary evolution faster than the disk evolution), can suffer significant interactions with a a circumbinary disk (CBD).  Hot accretion flows (``Hot AF'') will be generated when shearing and internal friction dominates over disk cooling mechanisms. Thus, PTA binaries might exist in a transition region of mixed thin standard accretion, self-gravitating, or hot, puffy disk flows. Lines of constant gravitational-wave inspiral timescale ($T_{\rm GW}$) and orbital semi-major axis ($a$, assuming Keplerian dynamics) are shown for convenience.
    }
    \label{fig:disks}
\end{figure}

The talk by D'Orazio argued that CW sources resolved by PTAs will likely be in the regime of either a self-gravitating or hot accretion flow, or near a decoupling point 
(Fig.\,\ref{fig:disks}, see talk by D'Orazio). The PTA measurement of CW source parameters may allow theoretical models to make GW-parameter-specific predictions for emission signatures. \action{Having a clear theoretical expectation for the mapping of mass/mass ratio/period parameter space can serve to critically direct multi-messenger search efforts, and critically aid scientists when assigning association confidence of a particular host to a detected CW signal.} 


It is clear that currently, the community has no unilateral agreement on the relative confidence of any one EM signature. However, there were several that were discussed as promisingly unique to binary systems:

\noindent \textbf{Periodic or quasi-periodic flux variability, \emph{especially with lensing flares}} in the optical, X-ray, and radio were highlighted as a potentially robust diagnostic due to the uniqueness of expected flare light curves. Lensing flares will only occur in highly inclined systems, but are difficult to replicate with a single black hole \citep[][see talks by D'Orazio, Smith]{lensingflares1,lensingflares2}.
Other periodic emission (from tidal streams, Doppler boosting, etc.) may occur for a larger range of inclinations, but non-flaring quasi-periodic behaviors are not unique to binary systems.

\noindent {\bf Spatially resolved, high-frequency radio/sub-mm imaging of separate accretion flows in a MBHB} could also provide strong evidence and a direct route for tracking orbits. However, \action{the requirements for such high-resolution imaging are not yet met by any instruments, and would involve milli-Jansky sensitivities at $\sim 1 - 10\mu$s  resolutions (space-based baselines}; see \eg\ \citealt{dorazioloeb18,mmorbittracking}). These angular scales are also expected to represent the angular extent of the binary semi-major axes that dominate the PTA band \citep[][Fig. 11]{nano-15yr-binaryproperties}.  \action{Improved modelling of flux expectations for mid-THz accretion flows in binary systems, and cross-referencing MBHB population models with the capabilities of upcoming space-based interferometric missions, would allow us to determine the feasibility of this technique for MMA follow-up.} 






\subsection{Large Galaxy Catalogs}\label{sec:catalogs}
\begin{figure}
    \centering
    \includegraphics[width=0.8\columnwidth,trim=65mm 0mm 65mm 0mm,clip]{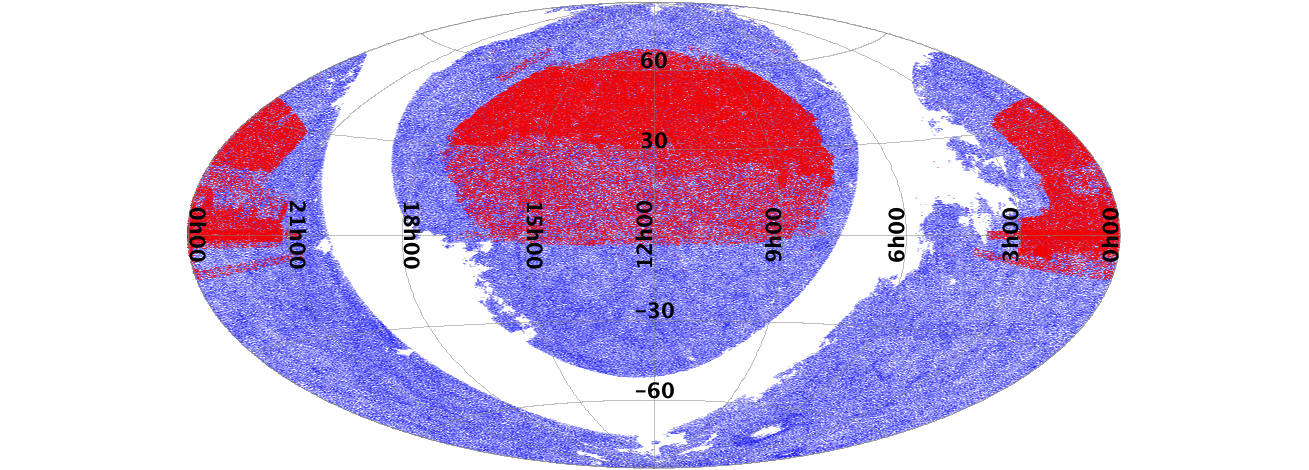}
    \caption{An R.A.-Dec. sky map showing the positions of the WISE x SuperCOSMOS photometric redshift catalog (blue), and of the sixteenth Data Release of the Sloan Digital Sky Survey Quasar Catalog (SDSS DR16Q, red). These surveys, and 2MASS, have already obtained spectroscopic and/or photometric redshifts for a significant portion of the targets likely to be of interest to PTAs, however may be missing the most distant possible hosts \citep{veronesi}, and larger errors in photometric redshift measurements may potentially lead to the erroneous omisson of a galaxy's inclusion in candidate host catalogs. Regions within a few degrees of the Galactic plane are incomplete in archives. Figure by N.~Veronesi.}
    \label{fig:skymap}
\end{figure}

It was broadly recognized that PTA localization regions (at the threshold-level of signal detection) will at first  be hundreds to thousands of square degrees, and it is likely this will improve gradually (see talk by Petrov). Therefore, 
archival imaging and spectroscopic catalogs should be leveraged to sub-select or rank host galaxies of interest. 
Large-field-of-view imaging optical or infrared surveys exist in most sky areas, although our sensitivity to hosts within $\pm\sim10^\circ$ of the Galactic Plane is incomplete (Fig.~\ref{fig:skymap}). 

Currently, it is not clear exactly what galaxy-scale properties can conclusively mark a binary host (see talk by Ruan). This is because the speed of binary inspiral, and relative timing of host morphological and spectroscopic merger signatures, are not completely settled. Numerous recent works \citep[\eg,][]{nevin+19,volonteri+20,host-properties23,charisi-500Mpc,bardati+24,horlaville+25} have explored the use of catalogs and simulations to identify the likely morphology of MBHB hosts.
While it has been suggested that a cut based on the ``MBHB mass and distance covariance region'' (\S\ref{sec:preliminaries}) may allow simple galaxy mass and distance filters, the conversion of MBHB to galaxy mass is prone to large errors, particularly for post-merger systems. In one evening discussion session, it was pointed out that while photometric redshifts are relatively complete for nearby, massive galaxy catalogs, they are prone to large errors, and this introduces a chance that the ``right galaxy'' might not be included in rank lists based on a mass/distance cut due to redshift inaccuracies. Spectroscopic surveys perform accurate redshift estimation, but \action{the mass/distance completeness of spectroscopic surveys to relevant galaxies as a function of sky position needs clarification.} 
SDSS has collected spectra of galaxies distributed over about one third of the sky, the vast majority of which are in the northern hemisphere, and approximately the same region is being observed by the Dark Energy Spectroscopic Instrument (DESI, \citealt{DESI+25}). In the next years, the 4-metre Multi-Object Spectroscopic Telescope (4MOST) will obtain spectra of galaxies in a large fraction of the southern sky. In particular, 4MOST's AGN survey will target the 80\%-90\% of the X-ray and mid-IR selected AGN samples over an area of 10,000 ${\rm deg}^2$, obtaining a total of up to one million spectra \citep{Merloni+19}. Combining these surveys, approximately the 70\% of the region of the sky with an absolute value of the galactic latitude of $|b|\geq10^\circ$ is expected to have spectral coverage in the near future.
IFU surveys have the most incomplete mass/distance/sky completeness, however can provide advanced kinematic and morphological parameterization that may be used to more accurately identify recent mergers \citep[\eg][]{horlaville+25}.



\action{It will be highly advantageous to understand what host-galaxy markers signify a likely MBHB host, and likewise to understand the completeness of galaxy catalogs to such markers.} As we note below in \S\ref{sec:cyberi}, it was agreed that \action{having a shared central repository of PTA events associated with value-added ranked candidate lists could streamline follow-up prioritization.} 
One significant question in the ``host galaxy morphology'' debate is the relative efficiency of post-merger galaxy relaxation vs.~that of MBHB formation/inspiral. Both of these are the subject of active research. It therefore seems likely that multiple metrics for host-galaxy candidate ranking may be published, providing separate host lists assessed using different metrics.

\begin{figure}
    \centering
    \includegraphics[width=1.0\columnwidth,trim=0mm 0mm 0mm 0mm,clip,]{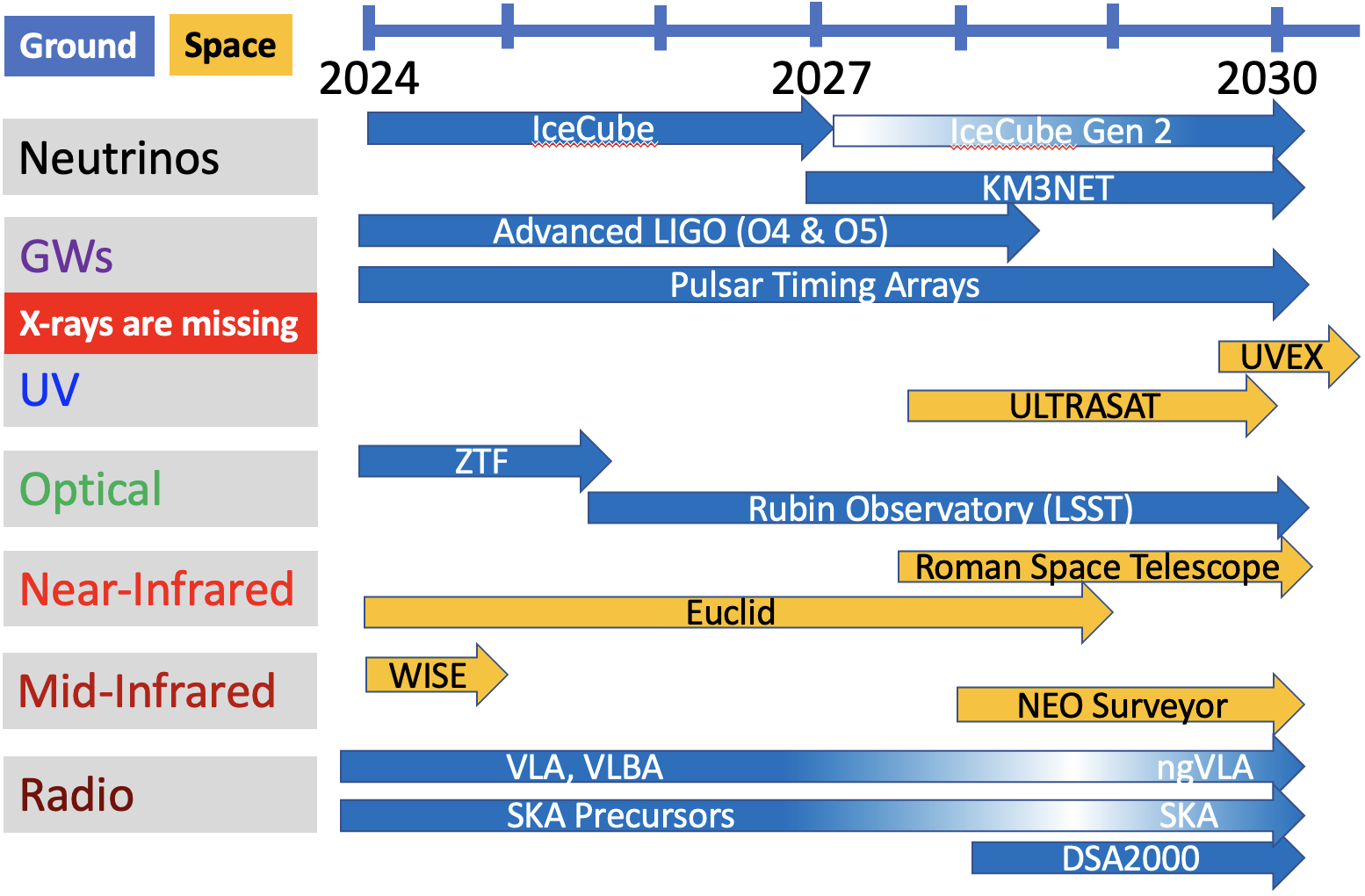}
    \caption{Large-field-of-view missions this decade (augmented based on summary talk by S.~Gezari). As of February 2025, the eRosita X-ray mission was paused and AXIS has not been approved, leaving a notable absence of sensitive, wide-field X-ray sensitivity.}
    \label{fig:missions}
\end{figure}

\subsection{Past and Future: Archives and Upcoming Observatories}
Session 5 focused on upcoming and present telescope resources for CW follow-up. It was noted that in designing PTA follow-up, rapid response is not typically required (unless a bursty or highly eccentric source is identified). The most beneficial elements will instead be large field-of-view or large-N-target capabilities, sensitivity to objects out to $z\sim0.4$, intermediate-cadence time-domain capabilities (providing sensitivity to 
$\sim$weeks-to-month-duration
lens flare events or disk flare events, \S\ref{sec:em}), and support for long-term (years-to-decades) return observations to capture properties over multiple binary orbits. 

As highlighted in \S\ref{sec:em}, MBHB follow-up is likely to build proof through observations across all wavelengths.
The summary of relevant ground- and space-based surveys shown in Figure~\ref{fig:missions} was presented in the Session 8 (Friday morning) discussion to assess missing capabilities.
High-cadence observations from LSST will be an excellent resource to identify periodic and lense-flare candidates. 
\action{X-Ray observations (identifying TDEs, self-Lensing, Fe K$\alpha$ features) can probe a wavelength range critical to circumbinary disk modeling, but as of yet there are no near-term new missions being launched.} PTA MBHBs are particularly promising X-ray sources because of their relatively high mass and low redshift, as long as the MBHB is accreting and visible as an X-ray AGN. The single and multi-epoch X-ray spectra of such sources will plausibly (if not likely) contain some useful information about the binary inclination, orbital phase, mass ratio, and possibly even spin (see talk in Session 5 by Malewicz). These EM-informed priors would be highly complementary to PTA GW detections, especially since some of this information, like spin, may not be immediately accessible with a PTA detection.

In addition, \action{
microarcsecond imaging at mid-THz wavelengths may provide a high-confidence direct imaging path (\S\ref{sec:em}),
but no current or planned mission exists with sufficient senstivitiy and resolution.}

Given the long timescales required, archival data will be of prominent importance, contributing valuable long-timescale measurements, though incomplete, that can improve our statistics and assessment of long-term behaviors.
Long-term monitoring data 
may need to be further mined from multiple disparate data archives for a large number of candidate objects. \action{Automation of archive mining for secondary EM indicators would be highly advantageous, and may be repurposed in part from existing LIGO/SCIMMA architectures.}



\subsection{Multi-band Gravitational-Wave Synergies}
While the meeting's focus was on nHz-frequency GWs, multi-band synergies were touched on in several sessions (talks by Steinle, Pardo, and numerous posters). 

Numerous studies were noted that use PTA results directly to inform predictions for LISA MBHB population predictions \citep{steinle+23, barausse+23}. PTAs and LISA will also (at high and intermediate-mass scales, respectively) together place novel constraints on galactic systems and jet physics traditionally observed via EMs, leading the way for multi-messenger constraints via astrophysical modeling.



Imaging-based gravitational wave detection may have comparable sensitivity to PTAs at the highest end of the PTA band, as highlighted in a talk by Pardo in Session 5. Thus, this may be a valuable independent detection method for candidate MBHBs at frequencies $-8\lesssim {\rm log}(f_{\rm gw})\lesssim -6$ (e.g. \citealt{astrometricgw}).

\begin{figure*}
    \centering
    \includegraphics[width=0.8\textwidth,trim=1mm 1mm 1mm 1mm,clip]{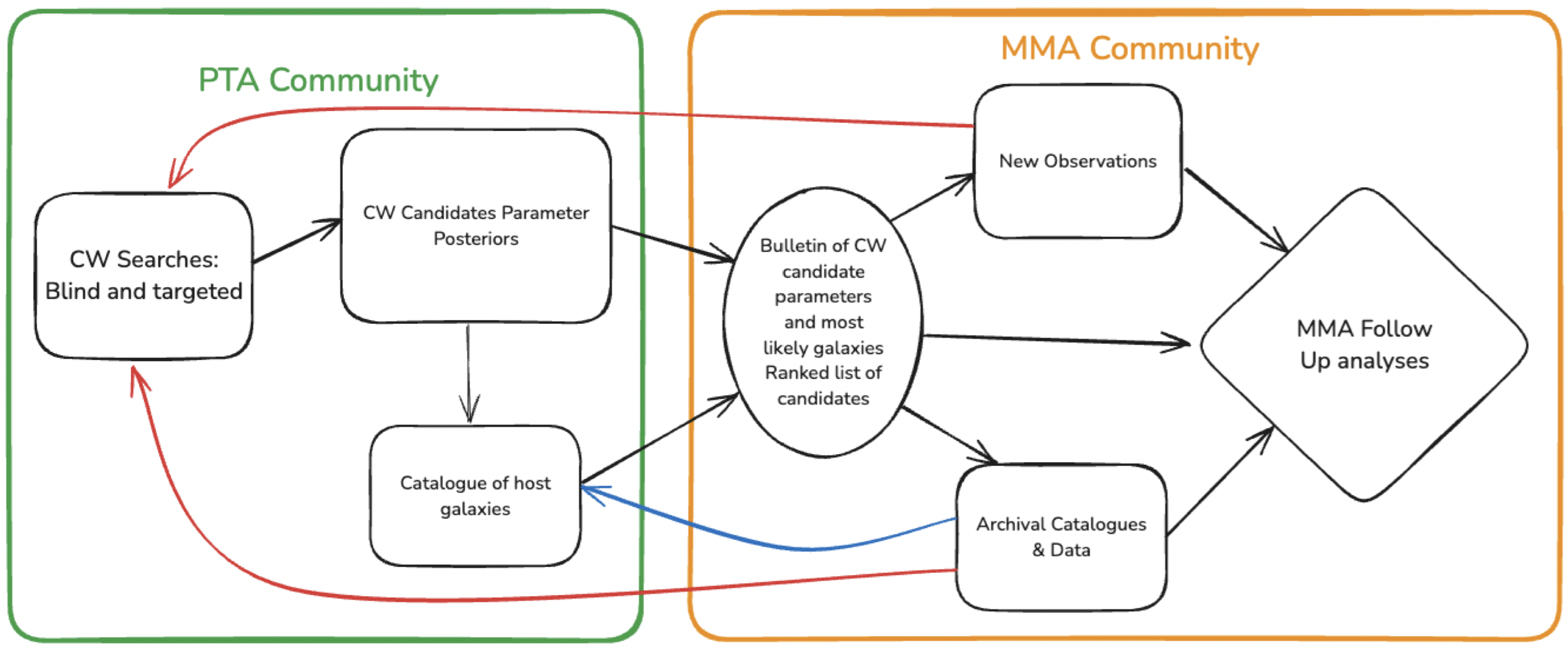}
    \caption{A sketch of the flow of information between the PTA community and MMA communities,  
    as described in \S\ref{sec:cyberi}. This figure was developed collaboratively by the Thursday evening communications discussion group.}
    \label{fig:cyberi}
\end{figure*}

\subsection{Cyber-infrastructure and Communications} \label{sec:cyberi}
An important practical aspect of conference discussion was cyber-infrastructure for use in PTA follow-up.
In Session 6 on Thursday morning, we discussed existing MMA cyber-infrastructure tools (used in LIGO follow-up), and the differences between the LIGO and PTA use-cases. 
A simple schematic developed in the evening discussion on this topic is shown in Figure~\ref{fig:cyberi}.

Discussion arose on which direction information should flow: will PTAs preferentially search for external events (periodic LSST targets, for instance) or will PTAs trigger electromagnetic searches? While LISA has ``signal verification binaries,'' and targeted detection searches are carried out regularly in LIGO for objects such as pulsars, in those cases target objects are well-supported examples of a given class. The MBHB and PTA case differs, because  as discussed in \S\ref{sec:em}, no EM emission detected to date provides a confident identification of an MBHB, and the PTA error region on the sky typically may contain many candidate EM systems and host galaxies (even after a mass/distance cut, Fig.~\ref{fig:banana}). It is possible that joint analysis of EM and CW data may result in improved detection confidence if the false alarm probability of EM signatures may be quantified, however joint EM/GW analysis is more likely to instead provide improved parameter estimation after a detection is demonstrated \citep[\eg][]{jointlikelihood}. 

Discussion thus progressed about the scenario that PTA-led detection will at first lead the pursuit of MBHBs multi-messenger association. 

Discussions converged on the fact that \action{a currently missing piece of infrastructure is a public database for discrete event communication, centrally created and maintained by one or more PTA groups, but ideally used by all PTAs under a common agreement.} Regardless of identification method (all-sky search, anisotropy, frequentist, Bayesian, IPTA or individual PTA), any released candidates should include all relevant gravitational wave parameters, such as the sky localization, orbital period, and estimated black hole masses, in addition to releasing formal statistical detection confidence indicators and parameter confidence regions, to guide the EM follow-up community in its decision making. 

If such infrastructure is developed, it was seen as \action{highly advantageous for PTAs to work in-house to develop value-added galaxy catalogs to accompany any detection release, given the broad availability of public photometric and spectroscopic catalogs}. 
As noted in \S\ref{sec:catalogs}, some groundwork for the generation of value-added catalogs is already underway.

Many groups are likely to pursue observation over extended (years to decade) time-scales to secure proof of association, therefore it was thought to be most realistic for extended EM follow-up will be self-organized by groups outside of PTA collaborations.
Tools are currently in development to easily access archival information on MBHB candidates previously published, including the tools BIGMAC \citep[see talk in Thursday Session 7][]{bigmac} and BOBcat \citep{jessica-thesis}, which will play a role in providing access to historical analysis on candidates in PTA error regions. It was stressed in the talk by Street that some organizations like NOIRLab and Rubin are beginning to offer access to data and computing resources that are housed in a single host location to allow remote interaction with large data sets. \action{A cyber-infrastructure ecosystem plan needs to be developed that identifies the unique needs of MBHBs and PTAs. It should leverage existing tools for observing coordination, data marshaling, and alert brokering.
This could be done in coordination with LISA cyber-infrastructure development, although the overlap needs deeper consideration.}



\subsection{Sociology}
Sociological considerations of multi-messenger and time-domain astronomy were a prominent topic of discussion throughout the conference. While there was no consensus opinion identified, multiple participants were vocal about the importance of recognizing scientifically rigorous work done by individuals within big teams, and rewarding not just ``first published'' but also ``scientific rigor,'' ``honest acquisition of data,'' ``equity,'' and ``willingness to contribute collaboratively.''



There was broad agreement that the contributions of all scientists (particularly early-career, e.g., students and postdocs) should be recognized. This could be done within existing MMA infrastructure, for instance through circulars that include reference to the data-proposal team, the analysis team, and the circular release team to track low-level contributions. Circulars can be linked to a citeable Digital Object Identifier (DOI) to allow an etiquette culture that involves direct tracking of contributions.

Central funding explicitly for MMA work, particularly for the important ``between the lines'' capabilities like cyber-infrastructure and the development of a code-savvy workforce, will be critical for maintaining coherence in the field. Similarly, there was some discussion about the important role of ``digital librarians'' in organizing data and providing resources to summarize, log, and disseminate an ecosystem of software and data resources.
\action{These capabilities could be explored through agency-based funding, or internship partnerships with private industry. This is not limited to PTAs/MBHB science, and explicit funding structures for MMA cyber-infrastructure need further development.}

It was noted that it will be sociologically difficult for the PTA community to release likely CW source locations and other information prior to publication, especially for the first detection. \action{Organization of CW source publication and public release requires explicit discussion in the PTA community over the coming years.}

\begin{figure}
    \centering
    \includegraphics[width=1.0\columnwidth,trim=0mm 0mm 0mm 0mm,clip]{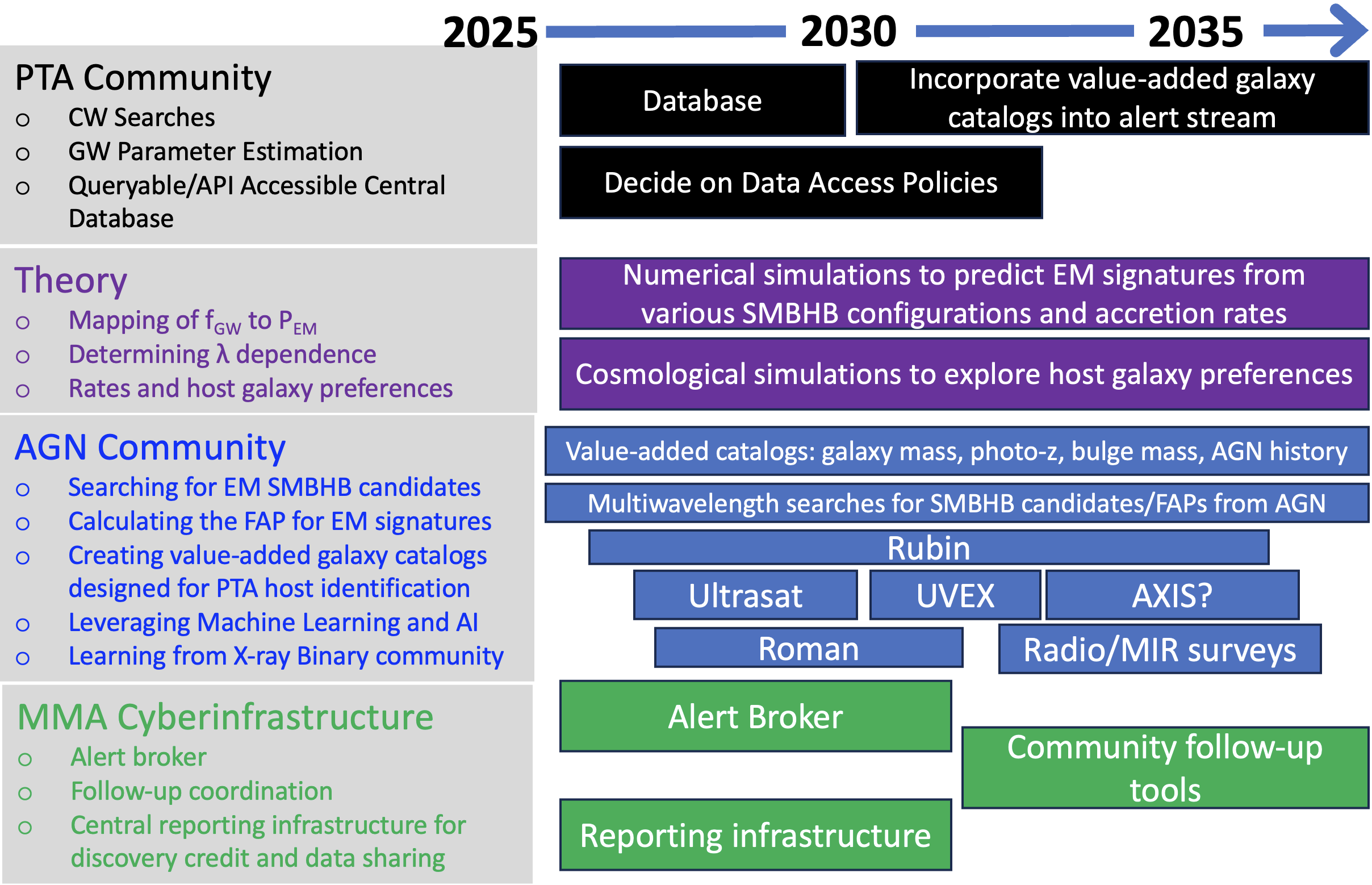}
    \caption{An end-of-conference roadmap as summarized by S.~Gezari, summarizing roles and some actions for relevant scientific communities.}
    \label{fig:enter-label}
\end{figure}

\section{Predictions for the Next Decade}
Returning now to our initial question, ``What is needed to take us from a PTA report of a CW detection to a confident MM identification?,'' 
here we summarize a possible sequence of events (with considerable uncertainties) to aid readers' thinking about planning and coordination of PTA follow-ups. It may be wrong, but we hope it catalyzes further discussion:
\begin{enumerate}
    \item Hints of a signal (sub-threshold significance) event will emerge in targeted CW searches, all-sky CW searches, or anisotropy searches. This may already be occurring (the excess in some sky regions in the MPTA data, sub-threshold peaks in continuous-wave searches, and/or peaks in the IPTA cross-power spectra).
    Given the large localization regions, low significance, and uncertainties in EM emissions, only small or partial attention may be paid to these early signals. 
    \item PTAs will eventually publish one or more at-threshold or above-threshold detections, supported by proof through a set of PTA statistical detection protocols.\footnote{The IPTA has an effort underway to create a formal set of CW detection protocol guidance.} It is possible that there will be disagreement in candidate parameterization between different search methods and PTAs. Thus, multiple likelihood regions with associated probabilities will likely be released.
    \item PTA collaborations may release value-added catalogs associated with their detection regions, containing ranked lists of candidate host galaxies based on large galaxy surveys and other resources.  Other groups may also release independent host galaxy ranked lists. 
    Other resources (like LSST brokers, online data archives, and catalogs that track MBHBs published in past literature like BIGMAC  and BOBcat, \citealt{bigmac,jessica-thesis}), will likely play an important role in ranking hosts of interest, although it is not yet clear whether handling of this data will be internal or external to PTA releases.
    \item Many different groups will organize observations that survey the PTA region or target value-added catalogs along different subselection criteria, gradually building up evidence that supports EM source/CW source associations within the region.
    \item 
    It is probable that a large, multi-wavelength set of observational links will be required to confidently tie the CW source to a host galaxy. This may take a long time, particularly because determining an EM-derived orbital period may be a critical point to link EM and CW emissions. Snapshot observations cannot currently provide definitive period information. Thus, EM observations may be required over partial or complete multi-year orbits. Significant ancillary science (particularly on circumbinary disk evolution, AGNs in post-merger systems, and related areas) will arise from these programs on the road to the first MBHB multi-messenger associations.
\end{enumerate}

\begin{figure}
    \centering
    \includegraphics[width=0.5\textwidth,trim= 30mm 40mm 60mm 100mm,clip]{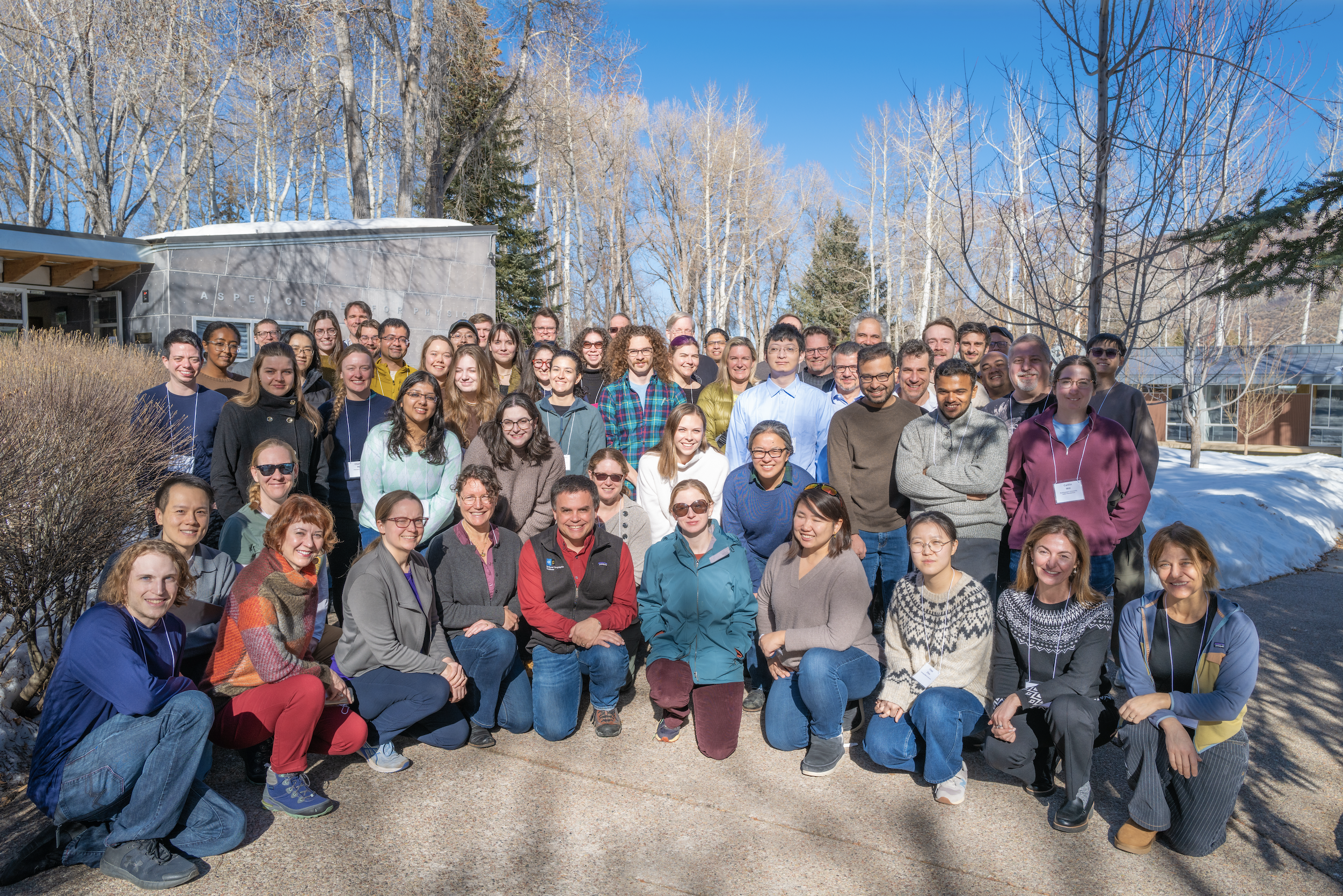}
    \caption{Conference participants thinking about black holes under a blue Colorado sky.}
    \label{fig:confphoto}
\end{figure}

\section{Acknowledgements}
This document was written by the authors and those who contributed its figures, but it presents contributions and ideas discussed by all participants listed in Section \ref{sec:participants}.

The conference took place at the Aspen Center for Physics, which is supported by National Science Foundation grant PHY-2210452. The SOC gratefully acknowledges significant support for this meeting by The Gordon and Betty Moore Foundation. We also gratefully acknowledge support from the Research Corporation for Science Advancement: nurturing teacher-scholars and innovative basic research in the physical sciences at American and Canadian colleges and universities since 1912. Further significant participant support for the conference was provided by the NANOGrav NSF Physics Frontier Center collaboration through NSF grant \#2020265. T.B. was supported in part by the grants from the Research Corporation for Science Advancement under award CS-SEED-2023-008 and the National Science Foundation under Grant No. AST-2307278. C.M.F.M.\ was supported in part by the National Science Foundation (NSF) under Grants No.\ NSF PHY-1748958, AST-2106552, NASA LPS 80NSSC24K0440, and in part by grant NSF PHY-2309135 to the Kavli Institute for Theoretical Physics (KITP). S.B.S.\ is supported in part by NSF Grant No.\ AST-2408649 and through a Sloan Fellowship.

\section{Alphabetical List of Participants}\label{sec:participants}
Thank you to the excellent discussions for all that attended this meeting. Most of the participants are pictured in Figure \ref{fig:confphoto}. This summary represents the SOC's take-aways from the meeting and plots generated by several participants, however we recognize the contributions to discussions of all participants listed below. The full participation list provided by ACP was:

{\small Nikita Agarwal (West Virginia University); Jillian Bellovary (CUNY - Queensborough Community College); Laura Blecha (University of Florida); Tamara Bogdanovic (Georgia Institute of Technology); Sukanta Bose (Washington State University); Adam Brazier (Cornell University); Michael Brotherton (University of Wyoming); Sarah Burke-Spolaor (West Virginia University); Yu-Ching Chen (Johns Hopkins University); Cecilia Chirenti (University of Maryland); Daniel J. D'Orazio (Space Telescope Science Institute); Jordy Davelaar (Princeton University); Tiziana Di Matteo (Carnegie Mellon University); Alberto Hernandez Diaz (Oregon State University); Alexander Dittmann (Institute for Advanced Study); Hannah Dykaar (University of Toronto); Kate Futrowsky (Georgia Institute of Technology); Suvi Gezari (Space Telescope Science Institute); Akshay Ghalsasi (Harvard University); Pratyasha Gitika (Swinburne University of Technology); Matthew Graham (California Institute of Technology); Kayhan Gultekin (University of Michigan); Daryl Haggard (McGill University); Jeffrey Hazboun (Oregon State University); Ziming Ji (Rochester Institute of Technology); Luke Kelley (University of California Berkeley); Matthew Kerr (Naval Research Laboratory); Anna Kyvernitaki-Synani (University of Crete); Shane Larson (Northwestern University); Anna-Malin Lemke (Universität Hamburg); Tingting Liu (West Virginia University); Julie Malewicz (Georgia Institute of Technology); Sean McWilliams (West Virginia University); Cole Miller (University of Maryland); Chiara Mingarelli (Yale University); Niana Mohammed (Pennsylvania State University); Beatrice Moreschi (Eleonora University of Milano Bicocca); Guatham Narayan (Univeristy of Illinois, Urbana Champaign); Rodrigo Nemmen (University of Sao Paulo); Scott Noble (NASA Goddard Space Flight Center); Jordan O'Kelley (West Virginia University); Feryal Ozel (Georgia Institute of Technology); Kris Pardo (University of Southern California); Polina Petrov (Vanderbilt University); Ryan Pfeifle (NASA Goddard Space Flight Center); Dimitrios Psaltis (Georgia Institute of Technology); Scott Ransom (National Radio Astronomy Observatory); Vikram Ravi (California Institute of Technology); John Ruan (Bishop’s University); Jessie Runnoe (Vanderbilt University); Pranav Satheesh (University of Florida); Jeremy Schnittman (NASA Goddard Space Flight Center); Krista Lynne Smith (Texas A\&M); Sloane Sirota (West Virginia University); Magdalena Siwek (Columbia University); Nathan Steinle (University of Manitoba); Rachel Street (Las Cumbres Observatory); Christopher Tiede (Niels Bohr Institute); Vishal Tiwari (Georgia Institute of Technology); Vivian U (University of California Irvine); Niccolò Veronesi (Washington State University); Sarah Vigeland (University of Wisconsin Milwaukee); Charlotte Ward (Princeton University); Caitlin Witt (Northwestern University); Kaiwen Zhang (University of Wyoming); Yihao Zhou (Carnegie Mellon University).}

\bibliography{main}{}
\bibliographystyle{aasjournal}

\end{document}